\documentclass[prd,showpacs,superscriptaddress,nofootinbib,floatfix,twocolumn]{revtex4-1} %notitlepage,
\usepackage[utf8]{inputenc}
\usepackage{amsmath,amssymb}
\usepackage{slashed}
\usepackage{graphicx}
\usepackage{subfigure}
\usepackage[colorlinks]{hyperref}
\usepackage[usenames,dvipsnames]{color}
\usepackage{mathrsfs}
\usepackage{color}
\usepackage{ulem}
\usepackage{bbm}               
\usepackage{xspace}				
\usepackage{epstopdf}
\usepackage{pgffor}		
\usepackage{array}
\usepackage{multirow}
\usepackage{graphicx}
\usepackage{microtype}

\newcommand{\uestc}{\affiliation{School of Physics, University of Electronic Science and
Technology of China, Chengdu 610054, China}}

\begin{document}

\title{Production of the weakly bound $T_{cc}^+$ state combining the coalescence model with its mass spectrum}
\author{Ao Wang}
\uestc

\author{Yu-Shan Ren}
\uestc

\author{Zhi Yang}\email{zhiyang@uestc.edu.cn}
\uestc

\date{\today}

%XXXXXXXXXXXXXXXXXXXXXXXXXXXXXXXXXX%
\begin{abstract}
We investigate the production of the doubly charmed molecular state $T_{cc}^+$ within the coalescence model. The formation probability is evaluated with the Wigner distribution function. In contrast to previous studies, which usually adopt a Gaussian wave function, we obtain the molecular wave function by numerically solving the coupled-channel Schrödinger equation with the one-boson-exchange potential fitted to the mass spectrum of $T_{cc}^+$. The resulting wave function is the universal form of a weakly bound $S$-wave molecule in momentum space (a Yukawa-type exponential in coordinate space), dressed by the dipole form factor of the OBE interaction, and is thus more suitable for such a state. As the input of the coalescence model, we use the transverse-momentum distributions of the constituent $D$ and $D^*$ mesons fitted to the ALICE and LHCb data, and also generated by PYTHIA for comparison. We compute the transverse-momentum distributions of $T_{cc}^+$ in $pp$ and Pb-Pb collisions. The two collision systems differ in which part of the wave function is probed: for the extended Pb--Pb source the universal pure-Lorentzian form suffices, whereas for the compact $pp$ source the short-distance part of the wave function, captured by the dipole form factor, becomes relevant. These results provide a theoretical reference for future experimental measurements of $T_{cc}^+$.
\end{abstract}

\maketitle
%XXXXXXXXXXXXXXXXXXXXXXXXXXXXXXXXXX% @ Begin

\section{Introduction}
Over the past two decades, hadron spectroscopy has been profoundly reshaped by the discovery of a growing family of states that defy the conventional quark-model classification. The experiments have compiled an extensive catalogue of exotic candidates, whose interpretation has become central to our understanding of the nonperturbative dynamics of quantum chromodynamics (QCD). Authoritative reviews of these states and of the theoretical pictures proposed for them can be found in Refs.~\cite{Chen:2022asf,Liu:2019zoy,Guo:2017jvc} and the references therein. 

In 2021, the LHCb Collaboration observed the first exotic state containing two heavy charm quarks and two light antiquarks ($cc\bar{u}\bar{d}$), called the $T_{cc}^+$ state, in the $D^0D^0\pi^+$ invariant-mass spectrum~\cite{LHCb:2021vvq,LHCb:2021auc}. The measured mass lies slightly below the $D^0D^{*+}$ threshold, the decay width is extremely narrow, and the spin-parity is established as $J^P=1^+$. These distinctive features make $T_{cc}^+$ an ideal candidate for exploring the formation laws of doubly heavy tetraquarks. Since its discovery, extensive theoretical studies have been devoted to its structure, binding mechanism, and decay properties. Within the molecular picture, $T_{cc}^+$ is generally interpreted as a loosely bound $S$-wave $DD^{*}$ state with isospin $I=0$~\cite{Feijoo:2021ppq}. Various approaches, including the one-boson-exchange (OBE) model~\cite{Li:2012ss,Molina:2010tx}, chiral unitary approach~\cite{Feijoo:2021ppq}, and lattice QCD~\cite{Padmanath:2022cvl,Lyu:2023xro}, have been employed and consistently support the near-threshold molecular nature of $T_{cc}^+$, reproducing its strong and electromagnetic decays~\cite{Meng:2021jnw}. The coupled-channel dynamics of the $DD^{*}$ system, as well as its bottom analogs, has been further analyzed in Refs.~\cite{Wang:2023ovj,Ren:2023pip}.

Compared with the mature studies on the energy spectrum and static structure of $T_{cc}^+$, investigations on its production and yield distribution are still in their infancy, yet they can reveal the intrinsic formation mechanism of doubly heavy tetraquarks from a dynamical perspective and serve as a crucial criterion to distinguish between the compact-tetraquark and hadronic-molecule pictures. For weakly bound hadronic molecules, the production mechanism has been studied extensively in the context of the $X(3872)$~\cite{Bignamini:2009sk,Artoisenet:2009wk,Albaladejo:2017blx}. Along this line, the production of hadronic molecules has also been investigated through $pp$ collisions~\cite{Guo:2014ppa,Guo:2014sca,Shi:2021hzm,Hua:2023zpa} and in semi-inclusive electroproduction~\cite{Shi:2022ipx}. The production of $T_{cc}^+$ has been investigated in the hadron-gas model~\cite{Abreu:2022lfy} and multiphase transport model~\cite{Hu:2021gdg}, while its isospin property has been probed via the Dalitz plot distribution~\cite{Shi:2022slq}. The coalescence model has also been applied to the production of exotic hadrons~\cite{ExHIC:2011say,Xu:2026rmz} and, in particular to $T_{cc}^+$~\cite{Yun:2022evm,Chen:2023xhd}, where the coalescence probability is encoded in a Gaussian Wigner function. The coalescence model with a Yukawa-type wave function has also been applied to other molecular states, such as the $\Xi N$ dibaryon~\cite{Wu:2023kqg}, where the Wigner density was nevertheless evaluated by expanding the wave function in a Gaussian basis. 

In this work we treat $T_{cc}^+$ as a hadronic molecule and employ the coalescence model to evaluate its production yield. We compute the Wigner distribution function directly from the molecular wave function obtained by solving the coupled-channel Schr\"odinger equation with the one-boson-exchange potential fitted to the mass spectrum of $T_{cc}^+$. The resulting wave function is the universal Lorentzian in momentum space (a Yukawa-type exponential in coordinate space) dressed by the dipole form factor of the OBE interaction, and its Wigner transform is evaluated analytically. This goes beyond both the Gaussian assumption and the Gaussian-basis expansion adopted in previous studies, and is more appropriate for such a weakly bound state. Combined with the $D^{(*)}$-meson spectra fitted to the experimental data, this treatment yields the transverse-momentum distributions of $T_{cc}^+$ in $pp$ and Pb-Pb collisions. The two collision systems are complementary. In heavy-ion collisions the abundant charm amplifies doubly charmed states and, with the large fireball, makes molecular coalescence efficient. This provides the most sensitive test of the molecular picture. In $pp$ collisions the charm multiplicity and source radius are much smaller, so the yield is strongly suppressed. Yet $T_{cc}^+$ was discovered in $pp$, making a reliable $pp$ prediction an essential baseline.

This paper is organized as follows. 
In Section~\ref{sec:formul}, we present the coalescence model, together with the production of the constituent $D^{(*)}$ mesons and the construction of the Wigner distribution function from the molecular wave function. The numerical results and discussions are presented in Section~\ref{sec:num}. Finally, a concise summary is given in Section~\ref{sec:sum}.

\section{Formulation} \label{sec:formul}
The production yield of the weakly bound molecule $T_{cc}^+$ is evaluated within the coalescence model, which requires two inputs: the phase-space distributions of the constituent $D$ and $D^*$ mesons, and the Wigner distribution function of the molecule. In the following, we first present the coalescence formula, then describe the production of the constituent $D^{(*)}$ mesons from the experimental data, and finally construct the Wigner distribution function from the molecular wave function.

\subsection{Coalescence model}
The coalescence model describes hadron production through the coalescence or recombination of its constituent particles~\cite{Gosset:1976cy,Greco:2003xt,Braun-Munzinger:2018hat}. Within this model, the differential yield of $T_{cc}^+$ is written as
\begin{widetext}
\begin{align} 
\frac{dN}{d^3\boldsymbol{p}} &= \sum_{i} g_i \int d^3\boldsymbol{x}_{1,i} d^3\boldsymbol{x}_{2,i} d^3\boldsymbol{p}_{1,i} d^3\boldsymbol{p}_{2,i} \,  f_{{1,i}}(\boldsymbol{x}_{1,i}, \boldsymbol{p}_{1,i}) f_{{2,i}}(\boldsymbol{x}_{2,i}, \boldsymbol{p}_{2,i}) W(\boldsymbol{x}_{1,i} - \boldsymbol{x}_{2,i}, \boldsymbol{k}_{i}) \,  \delta^3(\boldsymbol{p}_{T_{cc}} - \boldsymbol{p}_{1,i} - \boldsymbol{p}_{2,i}), 
\end{align}
\end{widetext}
where $\boldsymbol{x}_{1,i}$ and $\boldsymbol{x}_{2,i}$ are the spatial coordinates of the $D$ and $D^*$ mesons, respectively, while $\boldsymbol{p}_{1,i}$ and $\boldsymbol{p}_{2,i}$ denote their respective momenta, and $\boldsymbol{k}_{i} = (m_{2,i}\boldsymbol{p}_{1,i}-m_{1,i}\boldsymbol{p}_{2,i})/(m_{1,i}+m_{2,i})$ is the nonrelativistic Jacobi relative momentum of the pair, with $m_{1,i}$ and $m_{2,i}$ the masses of the two constituent mesons. The indices $i=1,2$ label the relevant charge channels. 
%In this equation, $\boldsymbol{x}_{D^+}$ and $\boldsymbol{p}_{D^+}$ are coordinate and momemtum of $D^+$, 
The functions $f_{{1,i}}(\boldsymbol{x}_{1,i}, \boldsymbol{p}_{1,i})$ and $f_{{2,i}}(\boldsymbol{x}_{2,i}, \boldsymbol{p}_{2,i})$ denote the phase-space distribution of the constituent $D$ and $D^*$ mesons, respectively. These functions can be normalized to their particle numbers, i.e. $\int d^3\boldsymbol{x}_{i}\, d^3\boldsymbol{p}_{i}\, f_{i}\bigl(\boldsymbol{x}_{i},\boldsymbol{p}_{i}\bigr) \equiv N_{i}$.
The Wigner distribution function of the composite particle $W(\boldsymbol{x},\boldsymbol{k})$ is normalized as $\int d^3\boldsymbol{x}\,d^3\boldsymbol{k}\, W(\boldsymbol{x},\boldsymbol{k}) = (2\pi)^3$. 
For the isospin-singlet $T_{cc}^+$, whose wave function reads $|T_{cc}^+\rangle=\tfrac1{\sqrt2}(|D^0D^{*+}\rangle-|D^+D^{*0}\rangle)$, the statistical factor $g_i=1/2$ denotes the statistical probability weight arising from the spin and isospin states when the two constituent particles coalesce into a composite state.

The phase-space distribution of each constituent meson factorizes as $f(\boldsymbol{x},\boldsymbol{p}) = \rho(\boldsymbol{x})\,dN/d^3\boldsymbol{p}$, where $dN/d^3\boldsymbol{p}$ represents the invariant momentum distribution of the particle. We assume the constituent mesons are emitted from a Gaussian spatial distribution
$\rho(\boldsymbol{x}) = \frac{1}{(\pi R^2)^{3/2}}\, e^{-\boldsymbol{x}^2/R^2}$. The source radius $R$ characterizes the spatial extent of the emission region (fireball) of the collision system. We take $R=4.08$ fm in Pb--Pb collisions at $0$--$10\%$ centrality, corresponding to a root-mean-square radius $\sqrt{3/2}\,R\approx 5$ fm, and $R=1.22$ fm in $pp$ collisions, corresponding to a root-mean-square radius $\sqrt{3/2}\,R\approx 1.5$ fm. The Pb--Pb value reflects the fireball radius of central heavy-ion collisions, whereas the $pp$ value is of the order of the proton radius. Both are of the same order of magnitude as the transverse emission radii inferred from hydrodynamical and Hanbury--Brown--Twiss analyses~\cite{HanburyBrown:1956bqd,HanburyBrown:1954amm,Lisa:2005dd}, although the latter probe the source of light hadrons rather than that of the charm mesons. The sensitivity of our results to the source radius is examined in Sec.~\ref{sec:num}. 

Taking the beam direction to be along the $z$-axis, the momentum distribution of $D^{(*)}$ mesons is independent of the azimuthal angle $\phi$. 
The longitudinal momentum differential satisfies $dp_{j,z} = E_j dy_j = \sqrt{m_{j,T}^2 + p_{j,z}^2} \, dy_j$, where $m_{j,T} = \sqrt{m_j^2 + p_{j,T}^2}$ is the transverse mass and $j = 1, 2$ refer to the $D$ and $D^*$ constituent mesons for the $D^0 D^{*+}$ and $D^{*0} D^+$ charge configurations, respectively.

The momentum distribution is expressed as $\frac{dN_j}{d^3\boldsymbol{p}_j} = \frac{1}{2\pi E_j} \frac{dN_j}{p_{j,T} \, dp_{j,T} \, dy_j}$. Introducing the relative-coordinate transformation $\boldsymbol{r} = \boldsymbol{x}_{1} - \boldsymbol{x}_{2},\quad \boldsymbol{R} = (\boldsymbol{x}_{1}+\boldsymbol{x}_{2})/2$ with the corresponding Jacobian determinant being unity, the spatial volume integration transforms into $d^3\boldsymbol{x}_{1} d^3\boldsymbol{x}_{2} = d^3\boldsymbol{r} \, d^3\boldsymbol{R}$. 
Since the integrand is independent of the center-of-mass coordinate $\boldsymbol{R}$, the integration over $\boldsymbol{R}$ yields the source-overlap factor
\begin{align}
S(\boldsymbol{r}) &= \int d^3\boldsymbol{R}\, \rho\bigl(\boldsymbol{R}+\tfrac{\boldsymbol{r}}{2}\bigr)\,\rho\bigl(\boldsymbol{R}-\tfrac{\boldsymbol{r}}{2}\bigr) = \frac{e^{-\boldsymbol{r}^2/(2R^2)}}{(2\pi R^2)^{3/2}}.
\end{align}
This isotropic Gaussian source neglects the boost-invariant longitudinal structure of the fireball and the correlation between the spatial and momentum rapidities. We thus obtain the final expression:
\begin{align}
\frac{dN}{d^3\boldsymbol{p}_{Tcc}}
&= \sum_{i} \frac{g_i}{(2\pi)^2} \int d^3\boldsymbol{r}_{i}\, d^3\boldsymbol{p}_{1,i}\, S(\boldsymbol{r}_{i}) \nonumber\\
&\quad\times
\frac{dN_{{1,i}}}{E_{{1,i}} p_{{1,i},T} dp_{{1,i},T} dy_{1,i}}\nonumber\\
&\quad\times
\frac{dN_{{2,i}}}{E_{{2,i}} p_{{2,i},T} dp_{{2,i},T} dy_{2,i}}\,
W\bigl(\boldsymbol{r}_{i},\, \boldsymbol{k}_{i}\bigr).
\label{eq:Tcc_coalescence}
\end{align}
This equation connects the final $T_{cc}$ yield to the momentum spectra and the spatial distribution of the constituent $D$ and $D^*$ mesons.

\subsection{Production of the constituent mesons}
As the input of the coalescence model, we use the transverse-momentum distributions of the constituent $D$ and $D^*$ mesons obtained by fitting the experimental data. For Pb--Pb collisions at $\sqrt{s_{NN}}=5.02$~TeV, we use the prompt $D^0$, $D^+$ and $D^{*+}$ spectra measured by the ALICE Collaboration in the $0$--$10\%$ most central collisions at midrapidity ($|y|<0.5$)~\cite{ALICE:2021rxa}. For $pp$ collisions at $\sqrt{s}=13$~TeV, we use the prompt $D^0$, $D^+$ and $D^{*+}$ spectra measured by the LHCb Collaboration in the forward region ($2<y<2.5$)~\cite{LHCb:2015swx}. Since the $D^{*0}$ spectrum is not measured, we take it to be equal to the $D^{*+}$ spectrum, following the treatment of Ref.~\cite{Yun:2022evm}.

The experimental spectra are available at a single rapidity interval. To evaluate the coalescence integral, which samples the constituent rapidities, we assume that the $D^{(*)}$ spectra are approximately independent of rapidity (boost-invariance) over the relevant window. For the midrapidity ALICE data this assumption is validated at the $1\%$ level~\cite{ALICE:2021rxa}. For the forward LHCb data, which are used here as the available high-statistics $pp$ measurement, the same approximation is adopted. Since we focus on the shape of the $T_{cc}^+$ $p_T$ distribution rather than on its absolute normalization, this approximation is adequate for the present purpose. 

The measured transverse-momentum distributions are parametrized with the two-piece form introduced in Ref.~\cite{Yun:2022evm}, whose explicit expression is given in Eq.~\eqref{eq:C1app} of the Appendix. The fitting parameters are collected in Table~\ref{tab:fitpar} of the Appendix. As discussed in Ref.~\cite{Yun:2022evm}, only a fraction of the measured prompt $D$ mesons participate in the coalescence at the formation point, because the excited charm states decaying into $D$ mesons are produced later. For comparison, we also use PYTHIA 8.3~\cite{Bierlich:2022pfr} to generate the corresponding distributions. For Pb--Pb collisions PYTHIA employs the Angantyr model~\cite{Bierlich:2018xfw}, which builds minimum-bias heavy-ion collisions from Glauber-initialized nucleon--nucleon collisions but does not contain a hot QCD medium. It therefore cannot account for the collective radial flow and the charm-quark energy loss that strongly reshape the measured $D$-meson spectra in Pb--Pb collisions, so its Pb--Pb results are expected to deviate from the ALICE data, especially at high $p_T$ where the data are suppressed by the in-medium energy loss. We follow Ref.~\cite{Yun:2022evm} and multiply the $D^0$ and $D^+$ spectra by the bare fraction $0.31$ to account for this feed-down, while the $D^*$ spectra are left uncorrected. This overall factor does not affect the shape of the computed $T_{cc}^+$ transverse-momentum distribution.

\subsection{From the wave function to the Wigner distribution}
The $T_{cc}^+$ is described as a loosely bound $S$-wave $DD^*$ molecule, and its wave function is obtained within the one-boson-exchange (OBE) model~\cite{Wang:2023ovj,Ren:2026wzt}. In this model the $DD^*$ interaction is generated by the exchange of the light pseudoscalar $\pi$ and vector $\rho$, $\omega$ mesons, and the coupling parameters are fixed by fitting the experimental $D^0D^0\pi^+$ lineshape of the $T_{cc}^+$ state~\cite{Wang:2023ovj}. In momentum space the effective potential reads
\begin{align}
\mathcal{V}({\boldsymbol{k}}_i,{\boldsymbol{k}}_f) &= \left[ V_{\pi}^{u}({\boldsymbol{q}}) + V_{\rho/\omega}^{t}({\boldsymbol{q}}) + V_{\rho/\omega}^{u}({\boldsymbol{q}}) \right] \nonumber\\
&\quad \times\left( \frac{\Lambda^2}{\Lambda^2 + {\boldsymbol{k}}_f^2} \right)^2
\left( \frac{\Lambda^2}{\Lambda^2 + {\boldsymbol{k}}_i^2} \right)^2,
\label{eq:potential}
\end{align}
where ${\boldsymbol{q}}={\boldsymbol{k}}_f-{\boldsymbol{k}}_i$ is the momentum transfer, the superscripts $u$ and $t$ label the $u$-channel and $t$-channel exchange diagrams, respectively, and the dipole form factors with cutoff $\Lambda$ regulate the ultraviolet behavior and encode the finite size of the interacting hadrons. The explicit expressions of the pion- and vector-meson-exchange potentials can be found in Ref.~\cite{Ren:2026wzt}. With the effective potential, the wave function of the $T_{cc}^+$ is obtained by the coupled-channel Schrödinger equation
\begin{align}
\mathcal{H}_{0}(\boldsymbol{k})\,\psi_{i}(\boldsymbol{k}) + \int_{0}^{+\infty} \frac{\boldsymbol{k}'^{2}\,d\boldsymbol{k}'}{(2\pi)^{3}} \, \mathcal{V}_{ij}(\boldsymbol{k},\boldsymbol{k}')\,\psi_{j}(\boldsymbol{k}') = E\,\psi_{i}(\boldsymbol{k}),
\label{eq:schrodinger}
\end{align}
where $\mathcal{H}_{0}(\boldsymbol{k})=\sqrt{m_{i,1}^{2}+\boldsymbol{k}^{2}}+\sqrt{m_{i,2}^{2}+\boldsymbol{k}^{2}}$ is the kinematic energy, with $m_{i,1}$ and $m_{i,2}$ the masses of the two constituent mesons in the $i$-th channel, $\mathcal{V}$ is the effective potential of Eq.~\eqref{eq:potential}, and the summation over the coupled channels $j$ is implied. Solving this equation numerically yields the wave function $\psi_{i}(\boldsymbol{k})$ of the $T_{cc}^+$. The resulting numerical wave functions, obtained in Ref.~\cite{Wang:2023ovj} with different choices of the cutoff, are almost identical. They are well described by the universal Lorentzian form of a weakly bound $S$-wave molecule~\cite{Artoisenet:2009wk}, dressed by the same dipole form factor that regularizes the OBE potential in Eq.~\eqref{eq:potential}:
\begin{align}
\psi(\boldsymbol{k}) &= \frac{\beta}{\boldsymbol{k}^2+\gamma^2}\left(\frac{\Lambda^2}{\boldsymbol{k}^2+\Lambda^2}\right)^2,
\label{eq:lorentzian}
\end{align}
where the first factor is the universal long-range wave function and the second factor is the dipole form factor that encodes the short-distance dynamics. The bare Lorentzian is the leading asymptotic form at long range (small momentum). The dipole factor, with the same cutoff $\Lambda$ as in the potential, well reproduces the correct suppression of numerical wave function at short-distance (large momentum). We have verified numerically that Eq.~\eqref{eq:lorentzian} reproduces the solution of Eq.~\eqref{eq:schrodinger} to better than $2\%$ over the whole relevant momentum range.
The normalization constant $\beta$ is fixed by $\int d^3k\,|\psi|^2=1$. In the limit $\Lambda\to\infty$ it reduces to the pure-Lorentzian value $\beta=\sqrt{\gamma}/\pi$. The binding momentum is $\gamma=\sqrt{2\mu B}$ with $\mu$ the reduced mass and $B$ the binding energy, which is determined in Ref.~\cite{Wang:2023ovj}. More specifically, $\gamma_{D^0D^{*+}}=27.6$ MeV and $\gamma_{D^+D^{*0}}=59.1$ MeV. It is instructive to note that in coordinate space this wave function is a superposition of the long-range Yukawa tail, $\psi(r)\propto e^{-\gamma r}/r$, and short-range terms $\propto e^{-\Lambda r}/r$ and $\propto e^{-\Lambda r}$. These short-range terms make $\psi(r)$ finite at the origin, while the long-range tail fixed by $\gamma$, is the universal part that sets the spatial size of the molecule. This long-range part is analogous to the situation for the deuteron, where the Hulth\'{e}n wave function $\psi(r)\propto(e^{-\alpha r}-e^{-\beta r})/r$ is adopted as a realistic bound-state wave function in coalescence calculations of light nuclei~\cite{Bellini:2020cbj,Mahlein:2023}. In contrast, most coalescence studies of exotic hadrons rely on a Gaussian wave function, which lacks the long-range exponential tail characteristic of a loosely bound state.
Because the $pp$ source is more compact than the Pb--Pb one, its coalescence integral probes higher relative momenta (shorter distances) and is therefore the more sensitive test of the short-distance part of the wave function. Quantitatively, the dipole form factor suppresses the integrated yield by about $15\%$ in $pp$, but changes it by only about $2\%$ in Pb--Pb.

%Wigner distribution is a quasi-probability distribution. It is defined as 
The formation probability of the $T_{cc}^+$ molecular state from its constituent $D$ and $D^*$ components is evaluated using the Wigner distribution function. In the relative coordinate space $\boldsymbol{r} = \boldsymbol{r}_1 - \boldsymbol{r}_2$ and relative momentum space $\boldsymbol{k}$, the Wigner function is defined as:
\begin{align}
W(\boldsymbol{r},\boldsymbol{k})
&=
\int d^3\boldsymbol{q}\,
\psi^*\left(\boldsymbol{k}+\frac{\boldsymbol{q}}{2}\right)
\psi\left(\boldsymbol{k}-\frac{\boldsymbol{q}}{2}\right)
e^{-i\boldsymbol{q}\cdot\boldsymbol{r}},
\label{eq:wigner_def}
\end{align}
%$\psi(r)$ is the wave function fo $T_{cc}$. 
where $\psi(\boldsymbol{k})$ is the momentum-space wave function of the $T_{cc}^+$ molecule. The exponential term accounts for the spatial Fourier transform, mapping spatial separation $\boldsymbol{r}$ to the momentum shift $\boldsymbol{q}$.
This formulation ensures that the near-threshold molecular nature of $T_{cc}^+$ and its spatial size are fully incorporated into the coalescence model during hadronization.

\begin{figure}[t]
\centering
\includegraphics[width=\columnwidth]{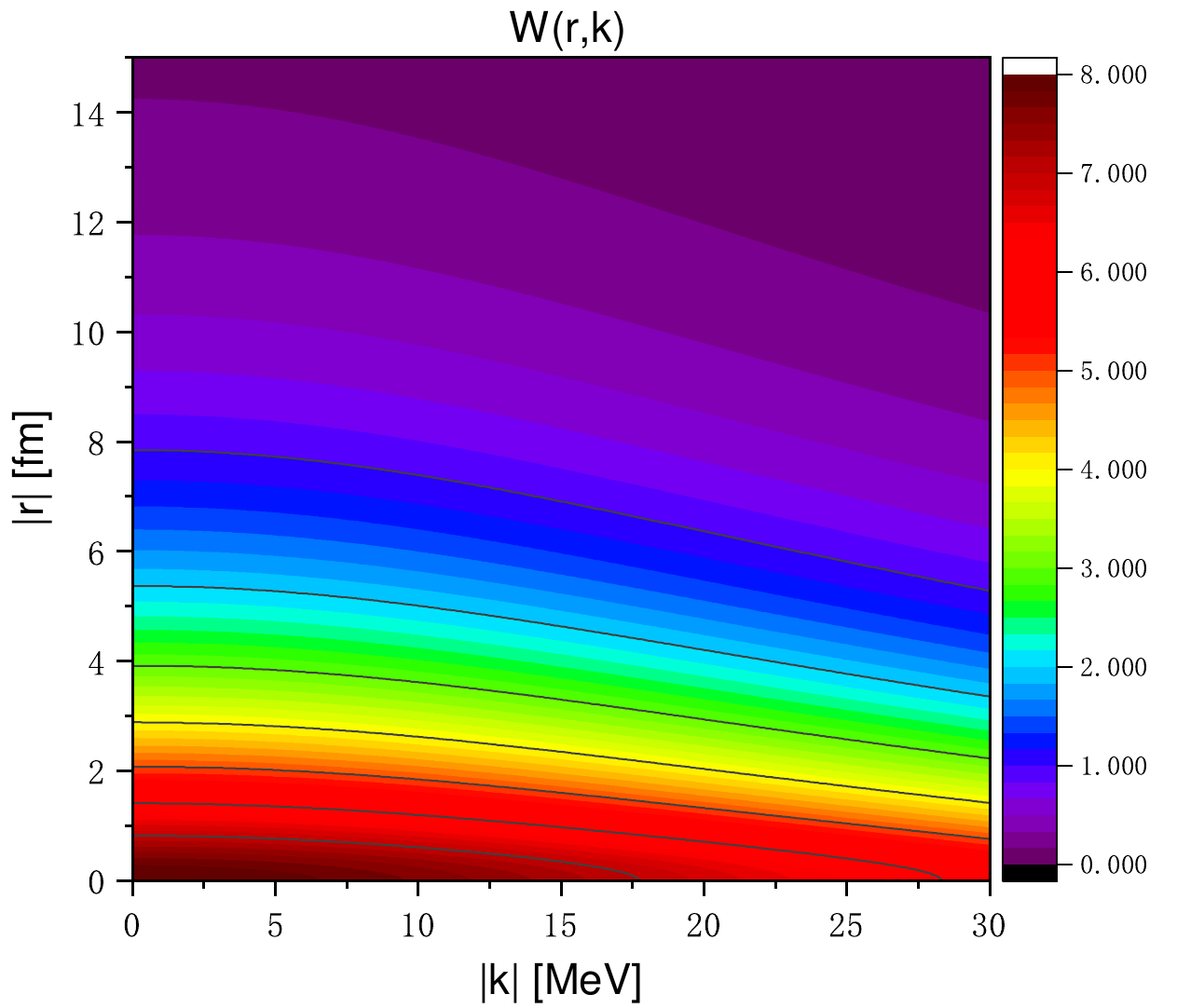}
\caption{\label{wigner_distribution}Wigner distribution \(W(\mathbf r,\mathbf k)\) of the \(T_{cc}^{+}\) molecular state as a function of the relative momentum \(|\mathbf k|\) and relative separation \(|\mathbf r|\), obtained from the Wigner transform of the momentum-space wave function of the $T_{cc}^+$ molecule.}
\end{figure}

For the wave function of Eq.~\eqref{eq:lorentzian}, it is convenient to use the partial-fraction decomposition
\begin{align}
\psi(\boldsymbol{k}) &= \frac{C_1}{\boldsymbol{k}^2+\gamma^2}+\frac{C_2}{\boldsymbol{k}^2+\Lambda^2}+\frac{C_3}{(\boldsymbol{k}^2+\Lambda^2)^2},
\label{eq:partialfrac}
\end{align}
with $C_1=\beta\Lambda^4/\delta^2$, $C_2=-\beta\Lambda^4/\delta^2$, $C_3=-\beta\Lambda^4/\delta$, and $\delta=\Lambda^2-\gamma^2$. Substituting this into Eq.~\eqref{eq:wigner_def} and employing the Feynman parametrization, the phase-space integration reduces to a single parametric integral,
\begin{align}
W(\boldsymbol{r},\boldsymbol{k}) &= 8\pi^2\!\int_{0}^{1}\!du\,\cos\!\bigl[2(1-2u)\boldsymbol{k}\cdot\boldsymbol{r}\bigr]\,
\mathcal{B}(u),
\label{eq:wigner_ff}
\end{align}
where $r\equiv|\boldsymbol{r}|$ and
\begin{align}
\mathcal{B}(u)&=C_1^2\,L(\Delta_\gamma)+C_2^2\,L(\Delta_\Lambda)+2C_1C_2\,L(\Delta)\nonumber\\
&\quad+2C_1C_3(1-u)\,D(\Delta)+2C_2C_3(1-u)\,D(\Delta_\Lambda)\nonumber\\
&\quad+C_3^2\,u(1-u)\,D_2(\Delta_\Lambda), \nonumber
\end{align}
with the elementary kernels
\begin{align}
L(\Delta)&=\frac{e^{-2r\Delta}}{\Delta},\qquad
D(\Delta)=e^{-2r\Delta}\!\left(\frac{r}{\Delta^2}+\frac{1}{2\Delta^3}\right),\nonumber\\
D_2(\Delta)&=e^{-2r\Delta}\!\left(\frac{r^2}{\Delta^3}+\frac{3r}{2\Delta^4}+\frac{3}{4\Delta^5}\right), \nonumber
\end{align}
and
\begin{align}
\Delta_\gamma&=\sqrt{\gamma^2+4u(1-u)\boldsymbol{k}^2},\quad
\Delta_\Lambda=\sqrt{\Lambda^2+4u(1-u)\boldsymbol{k}^2},\nonumber\\
\Delta&=\sqrt{u\gamma^2+(1-u)\Lambda^2+4u(1-u)\boldsymbol{k}^2}. \nonumber
\end{align}
In the limit $\Lambda\to\infty$ the form factor reduces to unity, and Eq.~\eqref{eq:wigner_ff} reduces to the pure-Lorentzian (universal) result
\begin{align}
W(\boldsymbol{r},\boldsymbol{k})
&= 8\gamma \int_{0}^{1} du
\frac{\cos\bigl[2(1-2u)\boldsymbol{k}\cdot\boldsymbol{r}\bigr]}
{\sqrt{\gamma^2+4u(1-u)\boldsymbol{k}^2}} \nonumber \\
&\quad \times \exp\left(-2|\boldsymbol{r}|\sqrt{\gamma^2+4u(1-u)\boldsymbol{k}^2}\right),
\label{eq:wigner_rp}
\end{align}
which is governed solely by the binding momentum $\gamma$. The Wigner function in Eq.~\eqref{eq:wigner_ff} results from reducing the three-dimensional integral of Eq.~\eqref{eq:wigner_def} to a single Feynman-parameter integral, which is then evaluated numerically.
In this way, the Wigner function smoothly bridges the spatial configuration of the $D$-$D^*$ pair with the momentum-space coalescence probability required for calculating the $T_{cc}^+$ production yield. In Fig.~\ref{wigner_distribution}, we show the Wigner distribution of the \(T_{cc}^{+}\) molecule. It is concentrated at small relative momentum set by the binding momentum $\gamma$ and extends in relative separation to the molecular size $1/\gamma$. The slow power-law falloff reflects the loosely bound and spatially extended structure of the state. In particular, at large relative momentum the Wigner function still falls only as a power law, inherited from the wave function and far less suppressed than for a Gaussian wave function. The dipole form factor makes this tail steeper than the pure-Lorentzian one above momentum $\Lambda$.

\section{Results and discussions}\label{sec:num}

With the phase-space distributions of the \(D\) and \(D^*\) mesons fitted to the experimental data (or generated by PYTHIA) and the Wigner function constructed from the molecular wave function as inputs, the coalescence model of Eq.~\eqref{eq:Tcc_coalescence} determines the three-dimensional momentum distribution \(dN/d^3p\) of the \(T_{cc}^{+}\). Integrating over the longitudinal momentum yields the transverse-momentum spectrum discussed below. This spectrum encodes two distinct pieces of physics: the collision source, which sets the constituent spectra, and the internal wave function, which enters through the Wigner function. 

\begin{figure}
\centering
\includegraphics[width=\columnwidth]{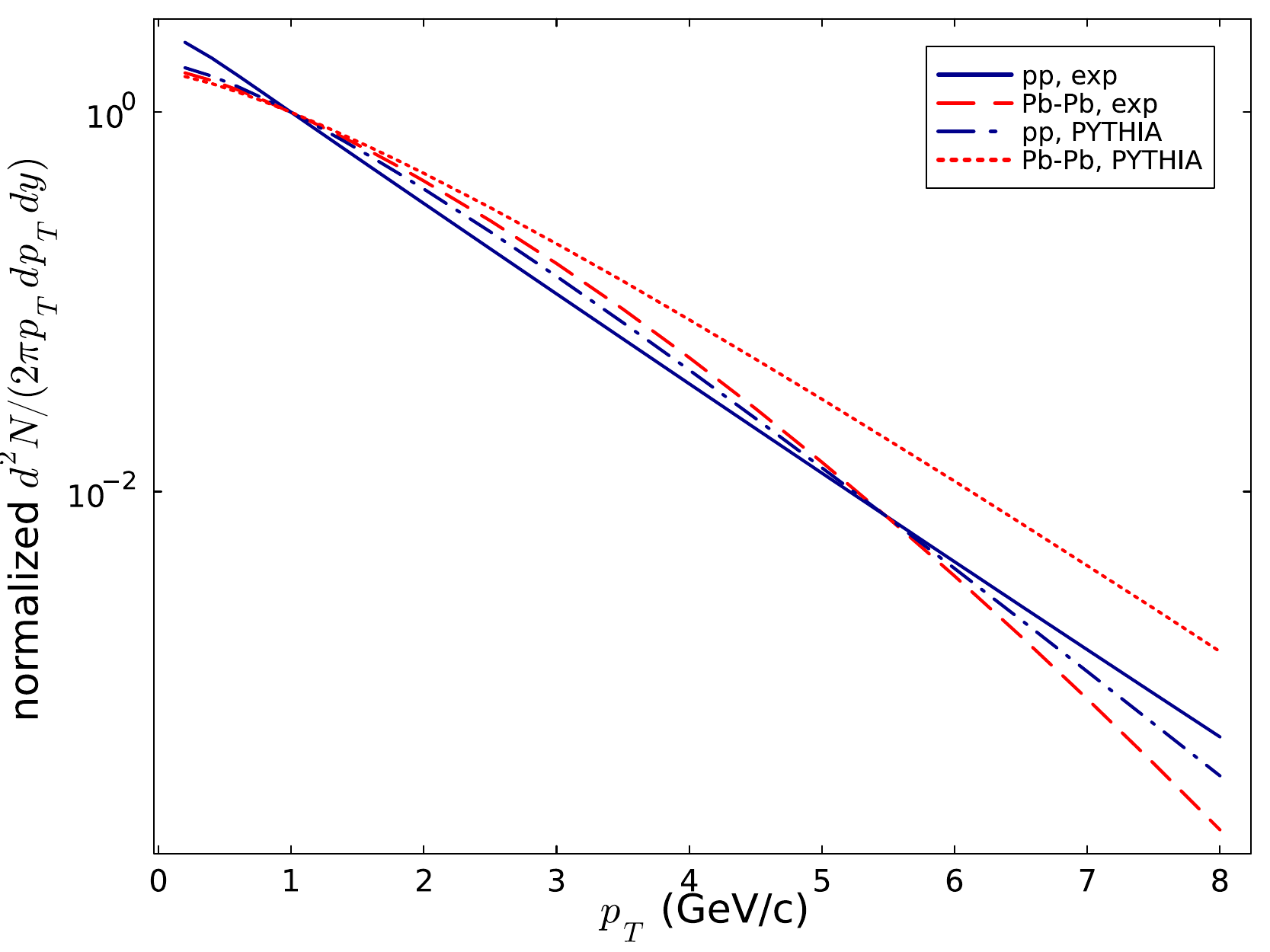}
\caption{Transverse-momentum distribution of \(T_{cc}^{+}\) in \(pp\) collisions at \(\sqrt{s}=13\) TeV and Pb–Pb collisions at \(\sqrt{s_{NN}}=5.02\) TeV, at \(|y|<0.5\) and normalized to unity at \(p_T=1\) GeV. The solid/dashed (dash-dotted/dotted) curves use the experimental (PYTHIA) constituent spectra for \(pp\)/Pb–Pb, respectively.}
\label{Tcc_pbpb_pp}
\end{figure}

The resulting transverse-momentum spectra in \(pp\) and Pb--Pb collisions are shown in Fig.~\ref{Tcc_pbpb_pp}, normalized to unity at \(p_T=1\) GeV. In both systems the yield decreases monotonically with \(p_T\), with the two spectra having comparable overall shapes. The \(pp\) spectrum falls somewhat more steeply at low and intermediate \(p_T\), while the Pb--Pb spectrum has a somewhat steeper tail at high \(p_T\). This difference reflects the different \(p_T\) dependence of the measured constituent spectra in the two collision systems. In particular, the steeper high-\(p_T\) tail of the Pb--Pb result reflects the stronger suppression of high-\(p_T\) \(D\) mesons in Pb--Pb collisions. The monotonic decrease can be understood as follows. At small \(p_T\) the two constituents are produced near-comoving, so that their relative momentum is well below the momentum scale \(\gamma\) set by the molecular wave function. As \(p_T\) increases, the pair must carry a larger total transverse momentum while the relative momentum still has to remain below this scale, so the accessible phase space shrinks and the spectrum falls off. The spectra obtained with the PYTHIA input are also shown in Fig.~\ref{Tcc_pbpb_pp}. In \(pp\) collisions the PYTHIA result is close to the experimental one over the whole \(p_T\) range, whereas in Pb--Pb collisions PYTHIA gives a substantially harder spectrum, especially at high \(p_T\), where it lies well above the experimental curve. This is expected, since the Angantyr model does not contain a hot QCD medium and thus cannot describe the charm-quark energy loss that suppresses the high-\(p_T\) \(D\) mesons measured by ALICE. 

%---it scales as \(1/R^3\) and is thus very sensitive to the source radius, and in particular for \(pp\) collisions the measured spectra are reported as cross sections while only a fraction of the prompt charm mesons participates in the coalescence
Since the absolute normalization of the computed yield is not fixed reliably, we present only the shape of the \(p_T\) distribution. The dominant uncertainty is the source overlap, which contributes a factor $S(\boldsymbol 0)\sim1/(2\pi R^2)^{3/2}\propto R^{-3}$: the $\sim\pm35\%$ spread on $R$ alone thus changes the absolute yield by roughly a factor of four. The yield also scales linearly with the bare prompt fraction ($\approx0.31$) that actually coalesces, and for $pp$ the cross-section-to-yield conversion adds further uncertainty. All these affect only the normalization, not the shape. We have verified that the shape is insensitive to the source radius: varying \(R\) by \(\pm35\%\) changes the normalized spectrum by only \(1\)--\(3\%\). Over the full range of plausible source radii ($R$ from $\sim1.2$ to $\sim7.7$ fm), the low-$p_T$ part remains stable while the high-$p_T$ tail varies by up to $6$--$7\%$, reflecting the truncation of the Wigner function by the source.

A central input of the calculation is the intrinsic momentum scale of the molecule. The binding momenta \(\gamma=27.6\) MeV and \(59.1\) MeV obtained for the \(D^0D^{*+}\) and \(D^+D^{*0}\) channels correspond to spatial extents \(1/\gamma\approx 7\) fm and \(3.3\) fm, underscoring the weakly bound, spatially extended character of the \(T_{cc}^{+}\). The corresponding momentum-space wave function is the universal Lorentzian dressed by the dipole form factor, whose power-law tail at large momentum, rather than the exponential tail of the Gaussian wave functions commonly assumed in the literature, is inherited by the Wigner function of Eq.~\eqref{eq:wigner_ff}. The coalescence probability is therefore less severely suppressed at large relative momentum, a feature that is essential for a reliable description of the long-tailed, near-threshold structure of the \(T_{cc}^{+}\) and that directly shapes the low-\(p_T\) yield.

\begin{figure}
\centering
\includegraphics[width=\columnwidth]{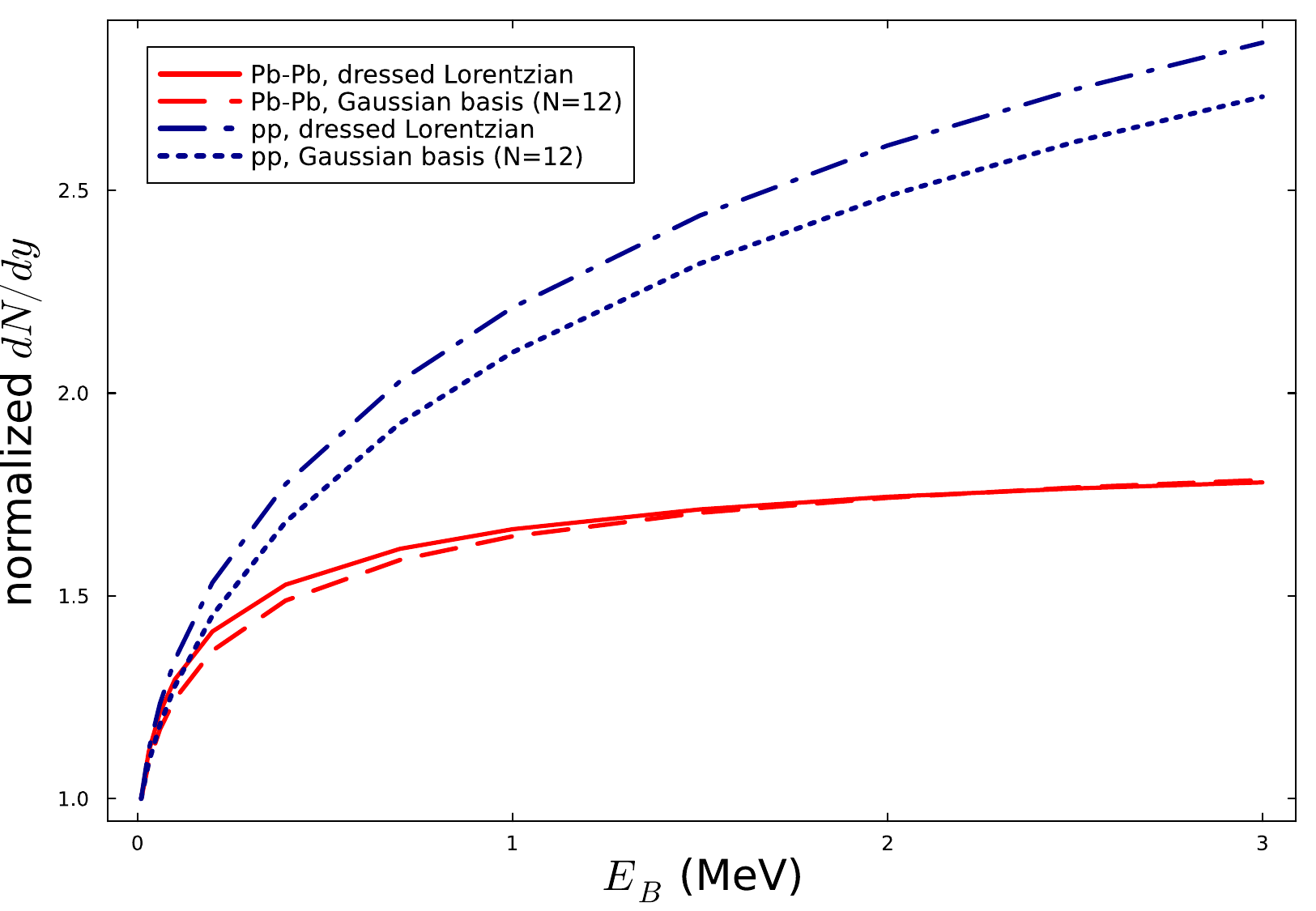}
\caption{Integrated yield $dN/dy$ of $T_{cc}^+$ in Pb--Pb collisions at $\sqrt{s_{NN}}=5.02$ TeV and $pp$ collisions at $\sqrt{s}=13$ TeV as a function of the binding energy $E_B$ (relative to the $D^0D^{*+}$ threshold), normalized to unity at $E_B=0.01$ MeV. The solid and dashed curves correspond to the dressed Lorentzian and the coherent 12-Gaussian basis expansion in Pb--Pb, while the dash-dotted and dotted curves correspond to the same in $pp$.}
\label{fig:binding}
\end{figure}

Since the binding energy of the $T_{cc}^+$ molecule is subject to theoretical uncertainty, it is instructive to examine how the production yield depends on it. In Fig.~\ref{fig:binding} we show the integrated yield $dN/dy$ as a function of the binding energy $E_B$ (measured with respect to the $D^0D^{*+}$ threshold), normalized to unity at the smallest value $E_B=0.01$ MeV, for the dressed Lorentzian wave function and for a coherent 12-Gaussian basis expansion of the same molecular wave function. The basis reproduces the dressed Lorentzian Wigner to better than $1\%$ at the physical binding energy, and the residual difference in the yield reflects the truncation of the finite Gaussian basis and shrinks as more terms are added. This illustrates the advantage of the dressed Lorentzian. It encodes the long-range tail in closed form and describes the molecular wave function faithfully, whereas a Gaussian basis needs many terms to approach it and still leaves a residual error. In both systems the yield rises and gradually flattens, the rise being somewhat steeper in $pp$ than in Pb--Pb and flattening only at larger $E_B$. The long-range tail of the molecular wave function leaves a clear imprint on the Wigner function, which retains a power-law tail in the relative momentum even for a spatially extended molecule; as a result the dressed Lorentzian yield already receives sizable large-relative-momentum contributions at small $E_B$.

It is also instructive to compare the $E_B$ dependence of our dressed Lorentzian result with the general expectation for the production of an $S$-wave threshold resonance. The factorization formula of Ref.~\cite{Artoisenet:2009wk} implies a long-distance factor proportional to $E_B^{1/2}$, so that the yield is suppressed only mildly as the binding energy decreases. In our calculation this long-distance factor, set by the long-range tail of the wave function, does scale as $\gamma\propto E_B^{1/2}$. However, the full $dN/dy$ is strongly flattened by the source overlap and by the presence of a second coupled channel. The small mass splitting between the two channels leaves the heavier one with a finite binding momentum, and hence an approximately $E_B$-independent contribution, even as $E_B\to0$. As a result the yield varies much more slowly with $E_B$ than $E_B^{1/2}$.

\section{Summary}\label{sec:sum}
In this work, we investigate the production of $T_{cc}$ in $pp$ and Pb-Pb collisions within the coalescence model.
In yield studies of $T_{cc}$ and other exotic hadronic states, the Wigner distribution function adopted in the coalescence model is usually assumed to follow a Gaussian form.
In the present work, however, the Wigner distribution is obtained via the Wigner transform from the momentum-space wave function of $T_{cc}$, which is well described by a Lorentzian form dressed by the dipole form factor. This Lorentzian form is the universal wave function of a weakly bound $S$-wave molecule discussed in Ref.~\cite{Artoisenet:2009wk}: it is determined solely by the binding momentum $\gamma$ and is independent of the detailed short-range dynamics. The long-range part of the resulting Wigner function is therefore universal, being fixed by $\gamma$, while the dipole form factor fixes the short-distance part. To the best of our knowledge, the direct analytic Wigner transform of the universal Lorentzian wave function has not been obtained before in the coalescence model of weakly bound $S$-wave molecules. 

Using the \(D\) and \(D^*\) momentum distributions fitted to the ALICE Pb--Pb and LHCb $pp$ data as input to the coalescence calculation, we obtain the transverse-momentum distributions of \(T_{cc}^{+}\) in pp and Pb–Pb collisions. We present the shape of the distributions, since the absolute normalization is not reliably fixed. In principle, once the full distribution of the source is known, the coalescence model would also yield the full distribution of the molecular state. However, our knowledge of the source distribution is still incomplete at present. A notable feature of our results is the mild fall-off at low \(p_T\), which is directly connected to the momentum distributions of the \(D\) and \(D^*\). Another distinctive feature is the dependence of the yield on the binding energy $E_B$: with the dressed Lorentzian wave function the yield flattens at large $E_B$ in Pb--Pb and continues to rise, though much more slowly than $E_B^{1/2}$, in $pp$. This dependence, which is insensitive to the overall normalization, provides an observable sensitive to the long-range tail of the molecular wave function. The present results provide a quantitative theoretical reference for future experimental measurements of the \(T_{cc}^{+}\) transverse-momentum distribution. Such measurements also provide valuable information on its underlying hadronic structure and offer a complementary approach to studying the molecular nature of this exotic state.

\bigskip
\appendix
\section{Parametrization of the constituent $D^{(*)}$ spectra}
\label{app:fit}

\begin{figure}[h]
\centering
\includegraphics[width=0.86\columnwidth]{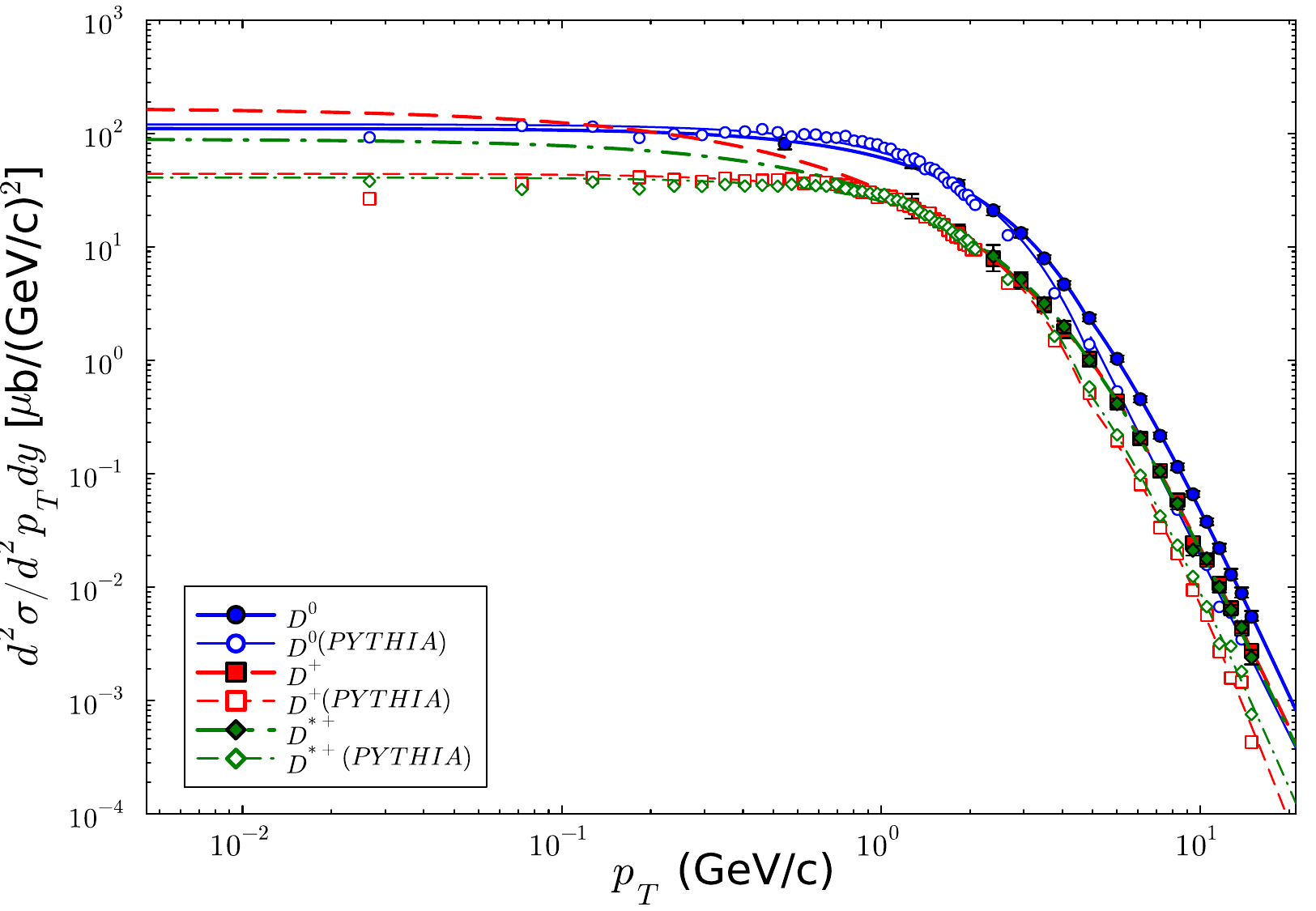}
\vspace{0.2em}
\includegraphics[width=0.86\columnwidth]{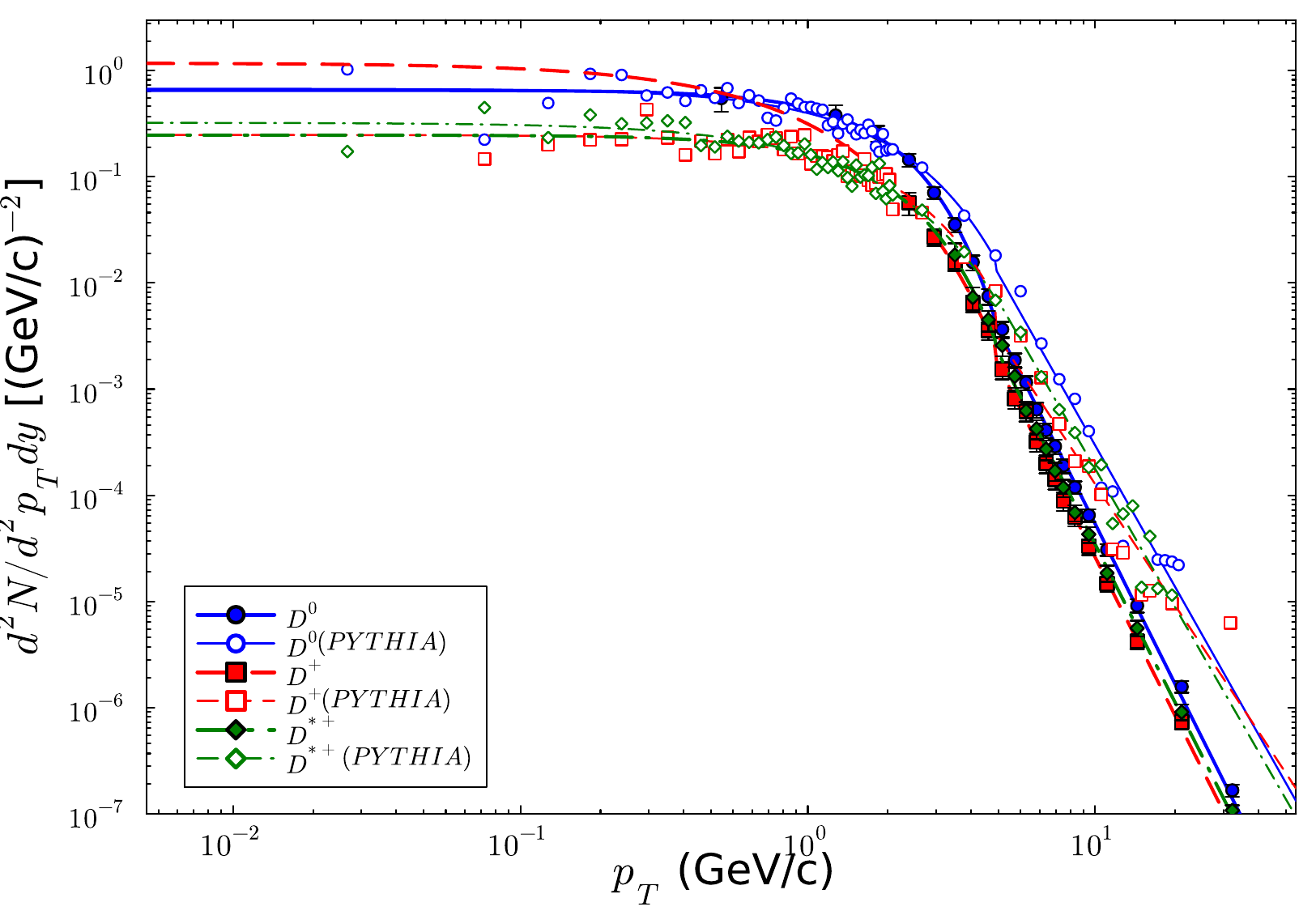}
\caption{The two-piece fits (lines) to the measured transverse-momentum distributions (points) of $D^0$, $D^+$ and $D^{*+}$ mesons in $pp$ collisions at $\sqrt{s}=13$ TeV and $2<y<2.5$ (top, LHCb data~\cite{LHCb:2015swx}) and in Pb--Pb collisions at $\sqrt{s_{NN}}=5.02$ TeV and $|y|<0.5$ (bottom, ALICE data~\cite{ALICE:2021rxa}), shown as the invariant yield. The PYTHIA-generated spectra and the corresponding two-piece fits are shown as open markers and thin lines, respectively, for comparison. The horizontal axis is logarithmic and extends down to a few MeV, the range of constituent momenta probed by the coalescence integral; below the measured range the curves are extrapolations of the two-piece fit.}
\label{fig:sourcefit}
\end{figure}

The measured transverse-momentum distributions of the constituent $D$ and $D^*$ mesons are fitted with the two-piece form
\begin{align}
\frac{d^2N}{d^2p_Tdy}
&= a\,e^{-b(p_T/p_1)^c} \qquad (p_T<p_c), \nonumber\\
&= \frac{d}{\bigl[1+(p_T/p_0)^2\bigr]^e} \qquad (p_T>p_c),
\label{eq:C1app}
\end{align}
with $p_1=1$~GeV. The fitting is performed on the invariant yield $d^2N/(d^2p_Tdy)$, which is obtained from the measured $dN/(dp_Tdy)$ by dividing by $2\pi p_T$. For Pb--Pb collisions we use the ALICE data in the $0$--$10\%$ most central collisions at $|y|<0.5$~\cite{ALICE:2021rxa}, and for $pp$ collisions the LHCb data at $2<y<2.5$~\cite{LHCb:2015swx}. The $D^{*0}$ spectrum is taken to be equal to the $D^{*+}$ spectrum. The resulting parameters are listed in Table~\ref{tab:fitpar}. Since the $D^{*+}$ spectrum in Pb--Pb collisions is only measured down to $p_T=3.25$ GeV, its low-$p_T$ exponential piece is not constrained by the data; in this case we fix the exponent $c=1.568$, taken from the upper-bound fit of Ref.~\cite{Yun:2022evm}, and fit the remaining parameters, which guarantees a physically finite low-$p_T$ extrapolation. For the other species the low-$p_T$ exponential piece is directly constrained by the data: the $pp$ $D^+$ spectrum starts at $p_T=1.25$ GeV (six data points below $p_c$) and the $pp$ $D^{*+}$ spectrum at $p_T=2.25$ GeV (four data points below $p_c$), so their exponent $c$ is left free in the fit. For comparison, the PYTHIA-generated $D^{(*)}$ spectra are also fitted with the same two-piece form and are shown together with the experimental data in Fig.~\ref{fig:sourcefit}.
\renewcommand{\arraystretch}{0.9}
\begin{table*}[h]
\centering
\caption{Parameters of the two-piece fit, Eq.~\eqref{eq:C1app}, of the constituent $D^{(*)}$ spectra. For Pb--Pb the ALICE data at $0$--$10\%$ centrality and $|y|<0.5$~\cite{ALICE:2021rxa} are used; for $pp$ the LHCb data at $2<y<2.5$~\cite{LHCb:2015swx} are used. The $D^{*0}$ spectrum is taken equal to the $D^{*+}$ spectrum. The exponent $c$ of $D^{*+}$ in Pb--Pb is fixed (see text). The PYTHIA-generated spectra are fitted with the same form, and for these the $D^{*0}$ and $D^{*+}$ are fitted separately. Note that the ALICE spectra are per-event yields, whereas the LHCb and PYTHIA spectra are cross sections, so the absolute values of $a$ and $d$ differ between the two systems by orders of magnitude. Only the shape is relevant for our purpose.}
\label{tab:fitpar}
\begin{tabular}{lcccccccc}
\hline\hline
species & system & $a$ & $b$ & $c$ & $d$ & $p_0$ (GeV) & $e$ & $p_c$ (GeV) \\
\hline
$D^0$    & $pp$    & $1.102\times10^{2}$  & $5.895\times10^{-1}$ & $1.258$ & $5.154\times10^{1}$  & $3.460$ & $3.124$ & $4.5$ \\
$D^+$    & $pp$    & $1.675\times10^{2}$  & $1.672$              & $7.438\times10^{-1}$ & $2.487\times10^{1}$  & $3.281$ & $3.010$ & $4.5$ \\
$D^{*+}$ & $pp$    & $8.888\times10^{1}$  & $1.120$              & $9.238\times10^{-1}$ & $2.314\times10^{1}$  & $3.346$ & $3.045$ & $4.5$ \\
$D^0$    & Pb--Pb  & $6.524\times10^{-1}$ & $3.900\times10^{-1}$ & $1.696$ & $2.024\times10^{7}$  & $7.328\times10^{-2}$ & $2.711$ & $4.5$ \\
$D^+$    & Pb--Pb  & $1.176$              & $1.319$              & $1.027$ & $4.050\times10^{6}$  & $8.192\times10^{-2}$ & $2.682$ & $4.5$ \\
$D^{*+}$ & Pb--Pb  & $2.467\times10^{-1}$ & $4.225\times10^{-1}$ & $1.568$ & $1.260\times10^{8}$  & $4.513\times10^{-2}$ & $2.678$ & $4.5$ \\
\hline
$D^0$    & $pp$ (PYTHIA)    & $1.746\times10^{1}$  & $8.026\times10^{-1}$ & $1.173$ & $1.253\times10^{8}$  & $1.209\times10^{-1}$ & $2.796$ & $4.5$ \\
$D^+$    & $pp$ (PYTHIA)    & $1.724\times10^{1}$  & $7.884\times10^{-1}$ & $1.186$ & $1.910$              & $4.412$ & $3.699$ & $4.5$ \\
$D^{*0}$ & $pp$ (PYTHIA)    & $1.361\times10^{1}$  & $7.247\times10^{-1}$ & $1.212$ & $4.789$              & $3.505$ & $3.446$ & $4.5$ \\
$D^{*+}$ & $pp$ (PYTHIA)    & $1.345\times10^{1}$  & $7.140\times10^{-1}$ & $1.217$ & $2.941\times10^{1}$  & $2.084$ & $2.951$ & $4.5$ \\
$D^0$    & Pb--Pb (PYTHIA)  & $1.456\times10^{5}$  & $6.035\times10^{-1}$ & $1.205$ & $2.333\times10^{13}$ & $3.374\times10^{-2}$ & $2.347$ & $4.5$ \\
$D^+$    & Pb--Pb (PYTHIA)  & $1.362\times10^{5}$  & $5.614\times10^{-1}$ & $1.244$ & $1.131\times10^{7}$  & $1.155$ & $2.833$ & $4.5$ \\
$D^{*0}$ & Pb--Pb (PYTHIA)  & $1.120\times10^{5}$  & $5.797\times10^{-1}$ & $1.191$ & $1.154\times10^{13}$ & $5.703\times10^{-2}$ & $2.517$ & $4.5$ \\
$D^{*+}$ & Pb--Pb (PYTHIA)  & $1.073\times10^{5}$  & $5.224\times10^{-1}$ & $1.261$ & $1.516\times10^{13}$ & $3.225\times10^{-2}$ & $2.291$ & $4.5$ \\
\hline\hline
\end{tabular}
\end{table*}
\renewcommand{\arraystretch}{1.0}

\begin{acknowledgments}
This work is partly supported by the National Natural Science Foundation of China (NSFC) under Grants Nos.~12275046 and 12547111.
\end{acknowledgments}

%=============== Refs ===============%  
\bibliography{ref.bib}

\end{document}